\documentclass[nobm]{epl}

\title{Band structure and atomic sum rules for x-ray dichroism}
\shorttitle{Band structure and x-ray dichroism}
\author{R. Benoist\inst{1} \thanks{Present address: 
Institut f\"{u}r Theoretische Physik C, Technische Hochschule, 
52074 Aachen, Germany.}\and P. Carra\inst{1} 
\and O. K. Andersen\inst{2}}
\institute{
  \inst{1} European Synchrotron Radiation Facility,
           B.P. 220, F-38043 Grenoble C\'{e}dex, France\\
  \inst{2} Max-Planck-Institut f\"{u}r Festk\"{o}rperforschung,
           Postfach 800665, D-70506 Stuttgart, Germany 
}

\begin{document}

\maketitle

\begin{abstract}
Corrections to the atomic orbital sum rule for circular magnetic x-ray
dichroism in solids are derived using orthonormal LMTOs as
a single-particle basis for electron band states.
\end{abstract}

Atomic physics affords a theory of x-ray dichroism 
by providing a set of sum rules which relate dichroic intensities, 
integrated over a finite
energy interval, to the ground-state expectation value of effective
one-electron operators \cite{Sta72vdL88,Tho92,Car93-1}. For circular
magnetic x-ray dichroism, that is the difference in absorption between
right- and left-circularly polarised photons in a system with a net
magnetisation, the effective operators coincide with spin and orbital
multipoles \cite{Car93-1,Car93-2}.

The spherical symmetry and the discreteness of the spectrum governing 
the atomic results do not hold for an atom in a solid, where the spin and
angular-momentum character selected by a particular x-ray transition is
spread out over a band of final states. This difference has hindered the
identification of a well-defined connection between the atomic sum rules and
band-structure calculations of magnetic x-ray dichroism \cite{Wu93}. 
Except in cases of strong electronic correlations, such calculations have been 
very successful in simulating experimental absorption spectra \cite{Guo94,Ebe96}.  
It therefore seems desirable to derive a band-structure formalism which
exhibits the atomic sum rules as the dominant term. This should be important
not only for the interpretation of x-ray dichroism in cases where current
density-functional band theory works, but also as a prerequisite for
understanding dichroism in strongly correlated materials. The current paper
is an attempt in this direction.

By leaving a localised hole, inner-shell photo absorption selects a specific
site in the solid, which we shall label by $R=0$. A local process is thus
expected to control the excitation, to leading order. Additional
contributions should emerge when the remaining sites in the lattice are
taken into account, that is, when electron delocalisation is included. In
this case, a minimal set of orthonormal linear muffin-tin orbitals (LMTOs)
provides a suitable single-particle basis \cite{An84,An94,An98}.

The macroscopic quantity of interest is the polarisation and
energy dependent extinction coefficient, 
$\kappa^{\makebox{\boldmath $\epsilon $}}(\omega )$  
\cite{Not1}. In a microscopic description, this is given by
\begin{displaymath}
\kappa ^{\makebox{\boldmath $\epsilon $}}(\omega )=
2\pi \left( c/\omega \right) ^{2}{\cal N\,}%
{\rm Im}f^{\makebox{\boldmath $\epsilon $}}(\omega )\,, 
\end{displaymath}
where $f^{\makebox{\boldmath $\epsilon $}}(\omega )$ 
stands for the forward scattering
amplitude, as determined by the ${\bf p\cdot A}$ coupling between x-rays and
electrons. Photon energy (in units of $\hbar $) and
polarisation are identified by $\omega $ and ${\makebox{\boldmath $\epsilon $}}$, 
respectively; 
${\cal N}$ denotes the number of excitable core electrons per unit volume. Only
electric-dipole transitions will be retained between spin-orbit coupled
inner orbitals, $\varphi _{\bar{n}\bar{l}\,\bar{j}\bar{m}_{j}}({\bf r,s}),$
localized around site $R=0,$ and spin-polarized, spin-orbit coupled band
states, $\psi _{k}\left( {\bf r,s}\right) $ \cite{Not4}. We consider only 
magnetic circular dichroism integrated over the two
partners $\left( \bar{j}=\bar{l}\pm \frac{1}{2}\right)$ of a given
spin-orbit split inner shell $\left( \bar{n}\bar{l}\right)$. For this, we have 
\begin{eqnarray}
&&
\int\nolimits_{\sim \,\varepsilon _{F}-\varepsilon _{\bar{n}\,\bar{l}\,
\bar{l}+\frac{1}{2}}}^{\sim \,\varepsilon _{c}-\varepsilon _{\bar{n}\,\bar{l}
\,\bar{l}-\frac{1}{2}}}\left[ \kappa ^{+}(\omega )-\kappa ^{-}(\omega )
\right] \,d\omega \;= \frac{(2\pi )^{3}}{3}{\cal N}\cos \theta \sum_{k}\langle \Psi
_{0}|a_{k}\,a_{k}^{\dagger }|\Psi _{0}\rangle \sum_{\bar{j}=\bar{l}\pm 
\frac{1}{2}}\sum_{\bar{m}_{j}=-\bar{j}}^{\bar{j}}  
\nonumber 
\\
&&
\left[ \left| \left\langle \psi _{k}\left| Q_{1\,1}\right| \varphi _{\bar{n}
\bar{l}\bar{j}\bar{m}_{j}}\right\rangle \right| ^{2}-\left| \left\langle
\psi _{k}\left| Q_{1\,-1}\right| \varphi _{\bar{n}\bar{l}\bar{j}\bar{m}_{j}}
\right\rangle \right| ^{2}\right] .  
\label{dic1}
\end{eqnarray}
Here, $\varepsilon _{F}-\varepsilon _{\bar{n}\,\bar{l}\,
\bar{l}\pm \frac{1}{2}}$ are the two threshold energies and 
$\varepsilon _{c}$ is a cut-off, positioned
far above the top of the valence band so that nothing would
change if $\varepsilon _{c}$ were increased by the spin-orbit splitting, 
$\varepsilon _{\bar{n}\,\bar{l}\,\bar{l}+\frac{1}{2}}-\varepsilon _{\bar{n}\,%
\bar{l}\,\bar{l}-\frac{1}{2}}$ of the inner level. The superscripts $\pm $
identify circular polarisations, and 
$Q_{1M}=eY_{1M}(\hat{{\bf r}})\,r $ is the electric dipole moment; 
$\theta $ is the angle between
photon wave vector and magnetization direction (orbital quantisation),
which we take along the $z$ axis; $\Psi _{0}$
denotes the ground state of the system, and $a_{k}^{\dagger }$ is a fermionic
creation operator for band states.

To evaluate $\left\langle \psi _{k}\left| Q_{1M}\right| \varphi _{%
\bar{n}\bar{l}\bar{j}\bar{m}_{j}}\right\rangle $, we use 
\begin{equation}
\varphi _{\bar{n}\bar{l}\bar{j}\bar{m}_{j}}({\bf r,s})=\sum_{\bar{m}\,\bar{m}
_{s}}C_{\bar{l}\,\bar{m};\frac{1}{2}\,\bar{m}_{s}}^{\bar{j}\,\bar{m}_{j}}
\varphi _{\bar{n}\bar{l}}(r)Y_{\bar{l}\bar{m}}(\hat{{\bf r}})\xi _{\bar{m}_{s}}
\left( {\bf s}\right) ,  
\label{inner}
\end{equation}
for the inner orbitals, that is a two-component function with the 
{\em same} radial dependence, $\varphi _{\bar{n}\bar{l}}(r),$ for 
$\bar{j}=\bar{l}-\frac{1}{2}$ and $\bar{l}+\frac{1}{2}.$ Notice that this is an 
excellent approximation as the radial probability densities, $4\pi
r^{2}[ f_{\kappa }( \varepsilon ,r) ^{2}+g_{\kappa }(\varepsilon ,r) ^{2}]$, 
for $\kappa {\rm =}l$ and $\kappa {\rm =}-l-1$ 
differ appreciably only close to the nucleus where they are small \cite{Note2};
here $f_{\kappa }(\varepsilon ,r) $ and $g_{\kappa }( \varepsilon ,r)$ are the 
solutions of the radial Dirac equations with $\kappa {\rm =}l,\,\varepsilon 
{\rm =}\varepsilon _{\bar{n}\,\bar{l}\,\bar{l}-\frac{1}{2}}$ and $\kappa 
{\rm =}-l-1,\,\varepsilon {\rm =}\varepsilon _{\bar{n}\,\bar{l}\,\bar{l}+
\frac{1}{2}}$ for the two partners, respectively. For $\psi _{k}$ we use an 
expansion in spin functions times spherical harmonics centered 
at the absorption site
\begin{equation}
\psi _{k}\left( {\bf r,s}\right) =\sum_{l\,m\,m_{s}}u_{lmm_{s},k}^{i}\,\phi
_{l}(r)\,Y_{lm}\left( {\bf \hat{r}}\right) \,\xi _{m_{s}}\left( {\bf s}
\right) .  
\label{valence}
\end{equation}
The inner orbitals are so localized that, in the region relevant to the integral 
$\left\langle \psi _{k}\left|
Q_{1M}\right| \varphi _{\bar{n}\bar{l}\bar{j}\bar{m}_{j}}\right\rangle$, 
the self-consistent field for a band electron is dominated by the Hartree
contribution, which is centrally symmetric and independent of $k$. As a consequence, 
the coefficients in expansion (\ref{valence}) factorise into normalization
constants, $u_{lmm_{s},k}^{i},$ and radial functions, $\phi _{l}(r),$ which
depend {\em only} on the magnitude of angular momentum about the absorption
site. Regarding relativistic effects, the argument given above for neglecting the 
$j=l\pm \frac{1}{2}$ dependence of the inner radial function holds also for
the radial functions of the band states. The exchange-correlation
potential {\em does} depend on the spin and possibly the orbital
character but, in the inner region, it amounts to $m_{s}$- and possibly $lm$%
-dependent shifts, which are {\em small} compared with the radial kinetic
energy, $\varepsilon _{k}-v\left( r\right) -l\left( l+1\right) /r^{2},$ in
that region. Compared with that kinetic energy, the band-energy range, 
$\varepsilon _{c}-\varepsilon _{F},$ is also small.

We now write the difference between the squared matrix elements for right-
and left-circularly polarized light as 
\[
\sqrt{2}\sum_{M=\pm 1}C_{1M;10}^{1M}\left\langle \psi _{k}\left|
Q_{1M}\right| \varphi _{\bar{n}\bar{l}\bar{j}\bar{m}_{j}}\right\rangle
\left\langle \varphi _{\bar{n}\bar{l}\bar{j}\bar{m}_{j}}\left| Q_{1M}^{\ast
}\right| \psi _{k}\right\rangle , 
\]
insert expressions (\ref{inner}) and (\ref{valence}), and apply the Wigner-Eckart
theorem. Notice that, owing to the sum over $\bar{j}$ in (\ref{dic1}), the
resulting expression is spin independent. The angular part is then recoupled
with use of the transformation \cite{VMK88}
\begin{eqnarray*}
&&
\frac{(-1)^{M+l-\bar{l}}}{2\bar{l}+1}\sum_{\bar{m}}C_{l\,m;1\,-M}^{\bar{l}
\bar{m}}\,C_{l^{\prime }m^{\prime };1\,-M^{\prime }}^{\bar{l}\,\bar{m}}= 
\\
&&
\sum_{l^{\prime \prime }\,m^{\prime \prime }}(-1)^{l^{\prime \prime
}-m^{\prime \prime }+l-m}\,C_{1\,M;\,1\,-M^{\prime }}^{l^{\prime \prime
}\,-m^{\prime \prime }}\,C_{l^{\prime }\,m^{\prime };\,l\,-m}^{l^{\prime
\prime }\,m^{\prime \prime }}\left\{ 
\begin{array}{ccc}
l & 1 & \bar{l} \\ 
1 & l^{\prime } & l^{\prime \prime }
\end{array}
\right\} ,
\end{eqnarray*}
with $l^{\prime \prime }=0,1,2$, corresponding to isotropic absorption,
circular, and linear dichroism, respectively. Hence, we obtain the result
\begin{equation}
\int\nolimits_{\sim \varepsilon _{F}-\varepsilon _{\bar{n}\,\bar{l}\,\bar{l}
+\frac{1}{2}}}^{\sim \varepsilon _{c}-\varepsilon _{\bar{n}\,\bar{l}\,\bar{l}
-\frac{1}{2}}}\left[ \kappa ^{+}(\omega )-\kappa ^{-}(\omega )\right]
\,d\omega = 
\pi ^{2}e^{2}{\cal N}\cos \theta \sum_{l=\bar{l}\pm 1}\frac{l-\bar{l}}{2l+1} 
R_{l1\bar{l}}\,\left\langle \Psi _{0}\left| L_{z}^{l}\right| \Psi
_{0}\right\rangle ,  
\label{main}
\end{equation}
with the radial integral over the inner region defined as
\begin{equation}
R_{l1\bar{l}}\equiv \left( \int_{0}^{\infty }\varphi _{\bar{n}\bar{l}}
\left( r\right) \,r\,\phi _{l}\left( r\right) \,r^{2}dr\right) ^{2},  
\label{radial}
\end{equation}
and $\int_{0}^{\infty }\varphi _{\bar{n}\bar{l}}\left( r\right)
^{2}r^{2}dr\equiv 1$. The operator $L_{z}^{l}$ is given by
\begin{equation}
L_{z}^{l}=\sum_{k}a_{k}a_{k}^{\dagger }\sum_{m}m\sum_{m_{s}}
\left| u_{lmm_{s},k}^{i}\right| ^{2}.  
\label{orbital}
\end{equation}
The individual normalisations of (\ref{radial}) and (\ref{orbital}) are irrelevant
when simulating the integrated dichroism (\ref{main}). This is because
normalization of the band states to unity in the solid merely fixes the
normalization of the product $u_{lmm_{s},k}^{i}\,\phi _{l}(r)$ in (\ref
{valence}). The usefulness of atomic sum-rules, however, stems from a
separation into an atomic factor which is independent of the
magnetisation direction, and a remainder which, for circular dichroism, 
is approximately the ground-state orbital angular momentum
of the excited atom. Suppose normalisations could be defined 
such that $u_{lmm_{s},k}^{i}$ were equal to the $R=0$ component of the
eigenvectors, $u_{Rlmm_{s},k}^{\perp }$, for the states
\begin{equation}
\psi _{k}\left( {\bf r,s}\right) =\sum_{Rlmm_{s}}\chi _{Rlm}^{\perp }
\left( {\bf r-R}\right) \xi _{m_{s}}\left( {\bf s}\right)  
u_{Rlmm_{s},k}^{\perp}\,,  
\label{band}
\end{equation}
in a representation of {\em orthonormal orbitals}. Then, from 
(\ref{main}) and (\ref{orbital}),
$\left\langle \Psi
_{0}\left| L_{z}^{l}\right| \Psi _{0}\right\rangle $ would be the
expectation value of the orbital angular momentum in the ground state, 
and the atomic sum rule would also hold in the solid.

Our use of LMTOs is motivated by the following features: 
they constitute a minimal basis whose
orthonormal representation and spherical-harmonic expansions about neighboring
sites are well known; their use in practical computations is well established\cite
{An84,An94,An98,An99}, even for systems with appreciable electronic
correlations \cite{LDAU}, and for systems with spin-orbit coupling and
spin-polarization \cite{relsp}; the simple formalism for the
orthonormal set, which we shall use below, was recently re-derived without
resort to the approximations of taking the interstitial kinetic energy equal
to zero and of dividing space into atomic spheres \cite{An94,An98,An99}.
Moreover, the LMTOs have recently been generalized to Nth-order MTOs 
spanning the states in a broad energy range, the accuracy and range increasing
with N, for a fixed basis-set size \cite{An99}. 

MTOs are derived from an MT-potential, i.e. a superposition of atom-centered
spherically-symmetric potential wells with ranges limited to about 0.7 times
the distance to the nearest neighbour \cite{An98}. In accordance with (\ref
{band}), we shall use MTOs derived from a spin- and orbital-independent
potential: 
$V\left( {\bf r}\right) \equiv \sum_{R}v_{R}\left( \left| {\bf r-R}\right| \right) $. 
To construct an LMTO, one first solves the
appropriate radial Schr\"{o}dinger or scalar-relativistic Dirac equation
for various energies in the band region, thus obtaining the radial functions $\phi
_{Rl}\left( \varepsilon ,r\right)$. Each of these is then multiplied by 
$Y_{lm}\left( \widehat{{\bf r}}\right) $ and, if we are using the atomic-spheres
approximation (ASA), truncated outside the atomic sphere. If not,
they are augmented continuously with tails; these are localized (screened) solutions
of the wave equation with the correct energies and are excluded from any inner
region \cite{An94,An98,An99}. As a result, we obtain the so-called
truncated or kinked partial waves, $\phi _{Rlm}( \varepsilon ,{\bf r-R})$. 
Now, the LMTO, 
\begin{equation}
\chi _{Rlm}\left( {\bf r-R}\right) \equiv \phi _{Rlm}\left( {\bf r-R}\right)  
+\sum_{R^{\prime }l^{\prime }m^{\prime }}\dot{\phi}_{R^{\prime }l^{\prime}m^{\prime }}
\left( {\bf r-R}^{\prime }\right) \,h_{R^{\prime }l^{\prime}m^{\prime },Rlm}\,,
\label{LMTO}
\end{equation}
centered at site $R$ and with spherical-harmonic character $lm,$ is defined
as the corresponding partial wave, taken at energy $\varepsilon _{\nu }$ 
at the centre of interest, plus a {\em smoothing cloud} of the first
energy derivatives, 
\begin{displaymath}
\dot{\phi}_{R^{\prime }l^{\prime }m^{\prime }}\left( {\bf r-R}^{\prime}\right) 
\equiv \left. \partial \phi _{R^{\prime }l^{\prime }m^{\prime}}
\left( \varepsilon ,{\bf r-R}^{\prime }\right) /\partial \varepsilon
\right| _{\varepsilon _{\nu }}, 
\end{displaymath}
of partial waves at their own and at neighboring sites. (Here, and in the
following, an omitted energy argument implies that $\varepsilon \equiv \varepsilon
_{\nu }$.) In (\ref{LMTO}), the expansion coefficients, $h$, form a
Hermitian matrix which is approximately the {\em band Hamiltonian}
with respect to $\varepsilon _{\nu },$ for the MT-potential used to generate
the LMTO set. Specifically, since the LMTO is smooth, we may operate with 
$(-\Delta +V-\varepsilon _{\nu })$ on each term in (\ref{LMTO}) to obtain 
$( -\Delta +V-\varepsilon ) | \phi ( \varepsilon )
\rangle =0.$ Energy differentiation then yields $( -\Delta
+V-\varepsilon ) | \dot{\phi}( \varepsilon )
\rangle =| \phi( \varepsilon ) \rangle$ and,
as a result, 
\begin{equation}
( -\Delta +V( {\bf r}) -\varepsilon _{\nu }) \,\chi
_{Rlm}( {\bf r-R}) =  
\sum_{R^{\prime }l^{\prime }m^{\prime }}\phi _{R^{\prime }l^{\prime}
m^{\prime }}\left( {\bf r-R}^{\prime }\right) \,h_{R^{\prime }
l^{\prime}m^{\prime },Rlm}\,.  
\label{Hop}
\end{equation}

As $\partial \phi ( \varepsilon ,{\bf r}) T( \varepsilon)/
\partial \varepsilon =\dot{\phi}\left( {\bf r}\right) T+\phi \left( 
{\bf r}\right) \dot{T}$, with $T$=1, changing the
energy-dependent normalization of a partial wave changes the shape of its
energy-derivative function by adding some amount of $\phi( {\bf r})$ to it. 
This in turn changes the shape of the
LMTOs via Eq. (\ref{LMTO}), but not the Hilbert space spanned by them. 
If each partial wave is normalized to one,  
we obtain a {\em nearly orthonormal} set since, in that case, 
the corresponding $\dot{\phi}_{Rlm}( {\bf r})$ is orthogonal to 
$\phi _{Rlm}( {\bf r})$, as
energy differentiation will reveal. Neglecting the overlap between
partial waves at different sites (ASA), or using L\"{o}wdin orthogonalisation 
\cite{An94,An98,An99}, one obtains 
\begin{displaymath}
\left\langle \phi _{Rlm}\mid \phi _{R^{\prime }l^{\prime }
m^{\prime}}\right\rangle =\delta _{RR^{\prime }}\delta _{ll^{\prime }}
\delta_{mm^{\prime }},\, \left\langle \phi _{Rlm}\mid \dot{\phi}_{R^{\prime }
l^{\prime }m^{\prime }}\right\rangle =0.
\end{displaymath}
Insertion into (\ref{Hop}) and (\ref{LMTO}) finally shows that, in the
nearly orthonormal representation, the LMTO Hamiltonian and the overlap matrices
are given by $\langle \chi | -\Delta +V-\varepsilon _{\nu }| \chi\rangle =h$ 
and $\langle \chi \mid \chi\rangle =1+hph$, respectively. 
(The off-diagonal elements of the matrix 
$p\equiv \langle \dot{\phi}\mid \dot{\phi}\rangle $ 
may be neglected.) The {\em truly orthonormal} set is therefore
\begin{equation}
\left| \chi ^{\perp }\right\rangle  =\left| \phi ^{\perp }\right\rangle
+\left| \dot{\phi}^{\perp }\right\rangle h^{\perp }=\left| \chi ^{\perp
}\right\rangle \left( 1+hph\right) ^{-\frac{1}{2}}
=\left( \left| \phi \right\rangle +\left| \dot{\phi}\right\rangle h\right)
\left( 1+hph\right) ^{-\frac{1}{2}},  
\label{OLMTO}
\end{equation}
where the expansion matrix,
\begin{equation}
h^{\perp }\equiv \left( 1+hph\right) ^{-\frac{1}{2}}\,h\,\left( 1+hph\right)
^{-\frac{1}{2}},  \label{h}
\end{equation}
is the band Hamiltonian 
without spin polarization and spin-orbit coupling, and where $\left| \phi
^{\perp }\right\rangle =\left| \phi \right\rangle 
\left( 1+hph\right)^{\frac{1}{2}}$ and $\left| \dot{\phi}^{\perp }
\right\rangle =\left( \left| 
\dot{\phi}\right\rangle -\left| \phi \right\rangle hp\right) \left(
1+hph\right) ^{\frac{1}{2}}.$

Next, we may work out the matrix elements of the exchange splitting and
spin-orbit coupling in the orthonormal representation (\ref{OLMTO}), add
them to $\varepsilon _{\nu }+h^{\perp },$ and diagonalize to find the
eigenvalues, $\varepsilon _{k},$ eigenvectors, $u_{Rlmm_{s},k}^{\perp },$
and band states (\ref{band}). Expanding the latter in spherical harmonics
about the excited site using (\ref{OLMTO}), we are finally able to identify 
the coefficients $u_{lmm_{s},k}^{i}$ in (\ref{valence}). 
(As usual, an omitted subscript $R$ implies that $R{\rm =}0.$) At first glance, 
it seems as if {\em two} radial integrals in (\ref{radial}) are needed: one
involving $\phi _{l}\left( r\right)$, as contributed by the head of the
LMTO, and the other involving $\dot{\phi}_{l}\left( r\right)$, as
contributed mainly by the tails of neighboring LMTOs. However we
observe that, when integrating the radial equation for the $l$-channel
outwards we may use the {\em same initial condition} for all energies.
Hence, we obtain an energy-derivative function, $\dot{\phi}_{l}^{i}(r)$, 
which is essentially excluded from the inner region, and whose
contribution to the integral (\ref{radial}) may therefore be neglected \cite
{An75}. Since this procedure amounts to choosing a particular
energy-dependent normalisation of the corresponding radial function, 
$\phi_{l}^{i}( \varepsilon ,r)$, the energy derivative function, 
$\dot{\phi}_{l}^{i}\left( r\right) ,$ must be a particular linear combination
of the $Y_{lm}$ projections of $\phi _{lm}( {\bf r}) $ and 
$\dot{\phi}_{lm}\left( {\bf r}\right) .$ These projections are
independent of $m$ in the ASA, where $\phi _{lm}\left( \varepsilon ,{\bf r} 
\right) =\phi _{l}\left( \varepsilon ,r\right) Y_{lm}\left( {\bf \hat{r}}%
\right) $, but only approximately independent when $\phi _{lm}\left( \varepsilon ,
{\bf r}\right) $ is a L\"{o}wdin orthonormalized kinked partial wave. In the
latter case, the $m$-dependence may be minimized through adjustment of the
screening \cite{An98,An99}; this dependence will be neglected in the present paper. 
Choosing to normalize $\phi _{lm}^{i}\left( \varepsilon ,{\bf r}\right) 
$ to one at $\varepsilon _{\nu },$ we can express the linear combination 
which does not contribute to the radial integral (\ref{radial}) as a
projection onto the orthonormal $\left( \phi ,\dot{\phi}\right)$ set 
\begin{equation}
\left| \dot{\phi}^{i}\right\rangle =\left| \dot{\phi}\right\rangle +\left|
\phi \right\rangle \left\langle \phi \mid \dot{\phi}^{i}\right\rangle \equiv
\left| \dot{\phi}\right\rangle +\left| \phi \right\rangle o^{i}.  
\label{oi}
\end{equation}
Here, $\langle \phi \mid \dot{\phi}^{i}\rangle \equiv o^{i}$ is a
matrix whose elements vanish unless $R{\rm =}R^{\prime }{\rm =}0$ and 
$l{\rm =}l^{\prime }=\bar{l}\pm 1,$ and whose off-diagonal elements are
neglected together with any $m$-dependence. We thus eliminate $\dot{\phi}%
_{lm}\left( {\bf r}\right) $ from (\ref{OLMTO}) and find
\begin{displaymath}
\left| \chi ^{\perp }\right\rangle =\left[ \left| \phi \right\rangle \left(
1-o^{i}h\right) +\left| \dot{\phi}^{i}\right\rangle h\right] \left(
1+hph\right) ^{-\frac{1}{2}}. 
\end{displaymath}
(One should keep in mind that $| \dot{\phi}^{i}\rangle=| \dot{\phi}\rangle$ 
unless $R{\rm =}0$ and $l=\bar{l}\pm 1$.)
Identification of (\ref{band}) with (\ref{valence}) yields
\begin{equation}
u_{k}^{i} =( 1-o^{i}h) ( 1+hph) ^{-\frac{1}{2}}u_{k}^{\perp }  
=( 1-o^{i}h^{\perp }-\frac{1}{2}h^{\perp }ph^{\perp }+...)
u_{k}^{\perp }.  
\label{ui}
\end{equation}
and, hence, the final result for use in (\ref{orbital}) is 
\begin{eqnarray}
&&
\left| u_{lmm_{s},k}^{i}\right| ^{2}=\left| u_{lmm_{s},k}^{\perp }\right|
^{2}-2\mathop{\rm Re} 
\left\{ u_{lmm_{s},k}^{\perp \ast }\,o_{l}^{i}\sum_{R^{\prime }
l^{\prime}m^{\prime }}h_{lm,R^{\prime }l^{\prime }m^{\prime }}^{\perp } 
u_{R^{\prime}l^{\prime }m^{\prime }m_{s},k}^{\perp }\right\}  
\nonumber
\\
&&
+\left| o_{l}^{i}\sum_{R^{\prime }l^{\prime}m^{\prime }}h_{lm,R^{\prime }
l^{\prime }m^{\prime }}^{\perp }\,u_{R^{\prime}l^{\prime }m^{\prime }
m_{s},k}^{\perp }\right|^{2}
\label{usq}
\\
&&
-\mathop{\rm Re}\left\{ u_{lmm_{s},k}^{\perp \ast }
\sum_{R^{\prime }l^{\prime }m^{\prime}}\left( h^{\perp }ph^{\perp }
\right) _{lm,R^{\prime }l^{\prime }m^{\prime}}\,u_{R^{\prime }l^{\prime }
m^{\prime }m_{s},k}^{\perp }\right\} +... 
\nonumber
\end{eqnarray}
The first term is contributed by LMTO heads only and gives the atomic sum
rule. Of the terms with $R^{\prime }\neq 0,$ the ones on the second line are
LMTO head-tail contributions, and those on the remaining lines are tail-tail
contributions. However, the sum of the terms on the second and third
lines may also contribute to the atomic sum rule, as they depend on 
$\varepsilon _{\nu }$. [This dependence is cancelled by the $\varepsilon
_{\nu }$ dependence of the radial integral brought about by
$\phi _{l}\left( \varepsilon _{\nu },r\right) $ in (\ref{radial}). Notice 
that to be able to neglect the $m$-dependence of
the valence orbital in the radial integral, we have chosen this orbital as
a partial wave rather than an LMTO, which has longer range.] 
To clarify this point, let us
assume that the spin-orbit interaction is smaller than the exchange 
splitting and that
the latter is fairly independent of $k.$ In this case, 
\begin{displaymath}
\sum_{R^{\prime }l^{\prime }m^{\prime }}h_{lm,R^{\prime }l^{\prime}
m^{\prime }}^{\perp }u_{R^{\prime }l^{\prime }m^{\prime }m_{s},k}^{\perp}
\approx \left( \epsilon _{k}-\varepsilon _{\nu }\right)
u_{lmm_{s},k}^{\perp }, 
\end{displaymath}
where $\epsilon _{k}$ is the (doubly degenerate) band without spin-orbit and
exchange couplings, and (\ref{usq}) reduces to 
\begin{eqnarray*}
&&
\left| u_{lmm_{s},k}^{i}\right| ^{2}\approx \left[ 1-\left( \epsilon
_{k}-\varepsilon _{\nu }\right) o_{l}^{i}\right] ^{2}\left|
u_{lmm_{s},k}^{\perp }\right| ^{2} 
\\
&&
-\mathop{\rm Re}
\left\{ u_{lmm_{s},k}^{\perp \ast }\left( \epsilon _{k}-\varepsilon _{\nu}\right) 
\sum_{R^{\prime }l^{\prime }m^{\prime }}\left( h^{\perp }p\right)
_{lm,R^{\prime }l^{\prime }m^{\prime }}\,u_{R^{\prime }l^{\prime }
m^{\prime},k}^{\perp }\right\} .
\end{eqnarray*}

We now realize that the {\em deviation} from the atomic sum rule is the
contribution to the integrated dichroism (\ref{main}) stemming from the 
second and further lines of (\ref{usq}), {\em after} they have been 
{\em minimized} with
respect to $\varepsilon _{\nu }$, that is, when $\varepsilon _{\nu }$ is
chosen as the centre of gravity of the unoccupied part of the $l$-projected
density of band states 
\begin{displaymath}
\varepsilon _{\nu l}=\frac{\sum_{k}\epsilon _{k}N_{k}^{l}}{\sum_{k}N_{k}^{l}}
,\quad N_{k}^{l}\equiv \left\langle \Psi _{0}\left| a_{k}a_{k}^{\dagger}
\right| \Psi _{0}\right\rangle \sum_{m\,m_{s}\,}\left| u_{lmm_{s},k}^{\perp}
\right| ^{2}.
\end{displaymath}
This choice of $\varepsilon _{\nu }$ is the one which also minimizes the
errors of the LMTO method. (When an $l$-independent $\varepsilon _{\nu }$ 
is used, the $o_{l}^{i}R_{l1\bar{l}}/( 2l+1)$-weighted average should be chosen).

To summarise: to estimate the accuracy of atomic sum rules for x-ray dichroism  
in solids, we have examined the problem of x-ray absorption by band electrons,
with emphasis on the interpretation of the total intensity of spectra
obtainable with circular polarisation in magnetic systems. Using an
orthonormal set of LMTOs, we have found corrections to the atomic results in the
form of energy moments of the band. Applications of the approach to
unpolarised and linear-dichroic spectra, together with a numerical
determination of the actual size of the corrections in specific cases will
be reported elsewhere.


\begin{thebibliography}{0}
\bibitem{Not1}  The absorption coefficient is defined by 
  $\mu ^{\makebox{\boldmath $\epsilon $}}(\omega)=
  (2\omega /c)\kappa ^{\makebox{\boldmath $\epsilon $}}(\omega ).$
\bibitem{Not4}  Translational symmetry is no prerequisite for our formalism.
  For states with translational symmetry, $k$ is a joint label
  for the band index and the Bloch vector.
\end{thebibliography}
\end{document}